\documentclass{kluwer}    

\newdisplay{guess}{Conjecture}
\usepackage{epsfig}
\begin{document}                                                                                   
\newcommand{\kap}[1]{Sect.\,\ref{#1}}
\newcommand{\sak}{\mbox{V4334 Sgr}}
\newcommand{\mem}[1]{\ensuremath{\mathrm{ #1}}}
\newcommand{\etal}{et~al.\,}
\newcommand{\kelv}{\ensuremath{\,\rm K}}
\newcommand{\emi}{\ensuremath{\mem{e}^\mem{-}}}
\newcommand{\teight}{\ensuremath{T_{\rm 8}}}
\newcommand{\msun}{\ensuremath{\, {\rm M}_\odot}}
\newcommand{\hevi}{\ensuremath{^{4}\mem{He}}}
\newcommand{\bel}{\ensuremath{^{11}\mem{B}}}
\newcommand{\ose}{\ensuremath{^{16}\mem{O}}}
\newcommand{\lsun}{\ensuremath{\, {\rm L}_\odot}}
\newcommand{\p}{\ensuremath{\mem{p}}}
\newcommand{\apleq}{\ensuremath{\stackrel{<}{_\sim}}}
\newcommand{\apgeq}{\ensuremath{\stackrel{>}{_\sim}}}
\newcommand{\mdot}{\ensuremath{\dot{M}}}
\newcommand{\kpc}{\ensuremath{\, \mathrm{kpc}}} 
\newcommand{\mzams}{\ensuremath{M_{\rm ZAMS}}}
\newcommand{\hedr}{\ensuremath{^{3}\mem{He}}}
\newcommand{\lisi}{\ensuremath{^{7}\mem{Li}}}
\newcommand{\besi}{\ensuremath{^{7}\mem{Be}}}
\newcommand{\jahre}{\ensuremath{\, \mathrm{yr}}}
\newcommand{\czw}{\ensuremath{^{12}\mem{C}}}


\include{abbr-engl}
\begin{article}
\begin{opening}         
\title{Modeling the evolution of Sakurai's object}
\author{Falk \surname{Herwig}}  
\runningauthor{Falk Herwig}
\runningtitle{Modeling the evolution}
\institute{Institut f\"ur Physik, Astrophysik, Universit\"at
  Potsdam, Potsdam, Germany \\
Institute for Physics and Astronomy, University of Victoria, Victoria
BC, Canada (fherwig@uvastro.phys.uvic.ca)}
\date{\today}

\begin{abstract}
Sakurai's object is a born again AGB
star of the \emph{very late thermal pulse} flavor. In this
contribution I will discuss new models of stellar evolution  and
nucleosynthesis models of this phase. Two most intriguing properties of
Sakurai's objects have so far not been understood theoretically: the 
peculiar chemical appearance, in particular the high lithium
abundance and the short time scale of only a few years on which the
transition from the dwarf configuration into the born again giant
appearance has occurred. A new nucleosynthesis mode of \emph{hot
  hydrogen-deficient \hedr\ burning} can explain the extraordinary
lithium abundance. During the thermal pulse \hedr\ is ingested from
the envelope together with the protons into the hot He-flash
convection zone. The first network calculations show that due to the
large \czw\ abundance protons are rather captured by carbon than
destroy newly formed \besi\ and ultimately \lisi. Moreover, the short
evolution time scale has been 
reproduced by making the assumption that the convective efficiency for
element mixing is smaller by two to three orders of magnitude than
predicted by the mixing-length theory. As a result the main energy
generation from fast convective proton capture will occur at a larger
mass coordinate, closer to the surface and the expansion to the giant
state is accelerated to a few years in excellent agreement with
Sakurai's behavior. This result represents an independent empirical
constraint on the poorly known efficiency of element mixing in
convective zones of the stellar interior.
\end{abstract}

\end{opening}           

\section{Introduction}  

Sakurai's object (\sak) has remarkable properties which
likely make it one of the most interesting stellar objects currently
investigated. Surrounded by a planetary nebula it has evolved
dramatically in recent years both in stellar parameters and in
chemical abundance pattern
\cite{pollacco:99,kerber:99,asplund:99a,duerbeck:00}.  The evidence
has been accumulating that the stellar evolution scenario for
\sak\ is that of a final He-shell flash \cite{iben:83a}
which occurs during the advanced central star of PNe (CSPN)
phase. However, the final He-flash might occur at different times
during the post-AGB evolution which results in important differences
in nuclear processing and mixing events. This sometimes confusing
detail will be recalled in \kap{sec:flavours}.

However, the observationally gathered information is more
numerous and detailed than can be accommodated by current stellar
evolution  and nucleosynthesis models. This is in particular true for
the abundance  
patterns recorded by Asplund \etal (\citeyear{asplund:99a}). The most
obvious feature is Sakurai's H-deficiency ($\mem{mass fraction} < 1\%$)
which is in accordance with
the final He-flash scenario. However, more puzzling is the high
abundance of lithium which exceeds the initial solar lithium
abundance by $0.5 \dots 1.0 \mem{dex}$. I will discuss a new
nucleosynthesis mode of lithium synthesis in \kap{sec:lithium-VLTP}.
 
Another unresolved property of the  evolution of \sak\ is the extremely
fast evolutionary pace from the dwarf configuration of a CSPN back to
the giant branch. 
All calculations of the born again phase published
so far are incompatible with $\tau_\mem{BA}(\sak) \simeq
5\jahre$\footnote{$\tau_\mem{BA} = t_\mem{AGB2} - t_\mem{TP}$,
$t_\mem{AGB2}$: time of arrival at AGB for 2nd time, $t_\mem{TP}$:
time of occurrence of final TP during post-AGB evolution.}. 
An analysis and solution to this
problem will be presented in \kap{sec:time} and \ref{sec:solution}.

\section{The different flavors of final He-flashes}
\label{sec:flavours}

Thermal pulses (TP) are a well known feature of the AGB evolution
\cite{iben:83b}. If a thermal pulse occurs during the CSPN phase the
star will return to the AGB \cite{schoenberner:79}. Depending on the
time at which the TP occurs during the post-AGB evolution two cases of
interest for \sak\ can be distinguished.

\subsection{The LTP case}
\label{sec:ltp}

If the TP occurs during the first part of the post-AGB
evolution (the \emph{crossing} of the HRD at constant luminosity) than
the H-burning shell is still active and the He-flash convection zone
in the intershell which forms during the He-flash is not   
able to extend mixing into the H-rich envelope. Previous calculations
of this \emph{late thermal pulse (LTP)} have shown that the H-rich
surface is preserved throughout the born again phase of evolution
\cite{schoenberner:79} including the second descent from the AGB
(e.g.\ Bl\"ocker \citeyear{bloecker:95b}). However,
recent calculations which included the effect of convective overshoot
have revealed that born again stars following a LTP can become
H-deficient after they have returned to the AGB
\cite{bloecker:00a,herwig:00a}. 

\begin{figure}[t]
\epsfxsize=8.8cm
\epsfbox{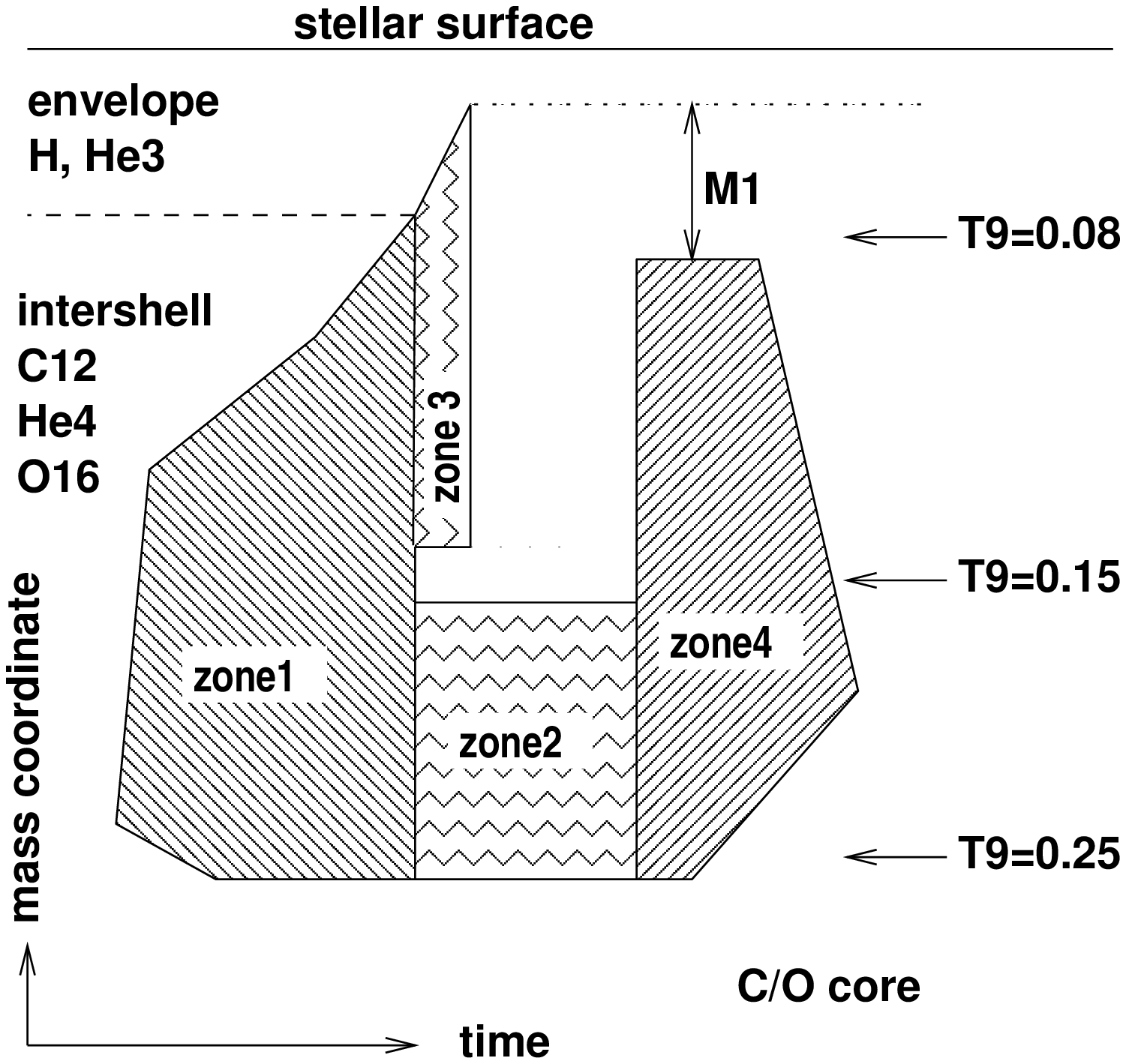}
\caption{\label{fig:Vs}  
Schematic (not to scale) of the time-evolution of convection zones in the top
$0.03\msun$ of a post-AGB star of $0.604\msun$ during a VLTP according
to the full evolutionary calculations by Herwig \etal (1999). All shaded
areas indicate convectively unstable zones. Approximate temperatures
are indicated  
in units of $10^9\kelv$ on the right side. Solid horizontal 
line at the top of the diagram: the stellar surface, dashed line: mass
coordinate of the envelope-intershell transition.}  
\end{figure}
The reason for the H-deficiency in the LTP case is a \emph{dredge-up}
event comparable to the third dredge-up on the AGB. Here, no
additional nucleosynthesis is involved. The surface abundance change
is only due to mixing processes. The time scale of the born again
phase of stellar evolution $\tau_\mem{BA}$ is of the
order of $100 \dots 200\jahre$ and therefore not in 
agreement with \sak\ (see \kap{sec:time}). Nevertheless, this evolutionary
scenario predicts a temporary increase of the lithium abundance as the
convective envelope 
forms and penetrates into the layer which contains the H-burning
profile (see Fig.\,6 in \mbox{Herwig, \citeyear{herwig:00a}}). It
reaches first the region which had experienced only partial pp-burning
and consequently formed a small \besi\ pocket. After the H-shell is
shut off by the LTP there remains enough time for the \besi\ pocket to
be transformed into a \lisi\ pocket by \emi\ capture. Once the bottom
of the convective envelope has passed through the former H-shell and
reaches the intershell, the surface abundance becomes
H-deficient. Clearly this scenario is not suitable for Sakurai's
object because it predicts the lithium rich phase while the star is
still H-normal. As the H-abundance  decreases by dilution so does
the lithium abundance.

It has been speculated whether the lithium abundance in \sak\ can be
due to the hot bottom burning (HBB) mechanism which is responsible for
the lithium production in massive AGB stars \cite{sackmann:92}. However, this
is not possible after a late TP.
The HBB requires the bottom of the envelope to reach down into the
H-shell. Immediately after a thermal pulse, however, H-burning is
inactive due to the expansion of the layer hosting the H-shell caused
by the He-flash. HBB can not operate without an active H-shell.

\subsection{The VLTP case}
\label{sec:vltp}

The evolutionary origin of \sak\ is a  \emph{very late thermal
  pulse} (VLTP). As the
name indicates the TP occurs later during the post-AGB evolution than in
the LTP case. The VLTP designates a TP which occurs after the CSPN has
evolved around the well known knee of post-AGB evolution and has started
its evolution towards the WD track at fading luminosities. At this
point the H-burning has stopped and the He-flash convection zone
reaches out into the H-rich envelope (see Fig.\,2 in Herwig
\citeyear{herwig:00a} and Fig.\,1). Unprocessed
material, most notably hydrogen 
(and \hedr\ if the progenitor AGB star has not experienced HBB) will
be ingested into the active He-flash convection zone. As the protons
are mixed inward their nuclear life time
against being captured by \czw\ will decrease as the temperature
increases. The peak energy by proton captures of \czw\ is released
where the nuclear time scale has decreased to the order of
the convective turn-over time scale. In the model by Herwig \etal
\citeyear{herwig:99c} the main proton capture location reaches a 
temperature of $\teight \sim 1.6$ which can be considered a typical
He-burning temperature.  

The additional energy release about $0.01\msun$ below the stellar
surface leads to a short period of enhanced convective instability
which is capable of rendering the entire outer layer homogeneous (see
Fig.\,5 in Herwig \etal\,\citeyear{herwig:99c}) and thus creating a
H-deficient post-AGB star by mixing \emph{and} burning processes. Special
effort has been made in order to achieve a consistent description of
the fast hydrogen burning at high temperatures. 

For the  calculation by Herwig \etal\cite*{herwig:99c} we developed a fully
coupled numerical scheme for convective nucleosynthesis. The equations 
of material transport (one diffusion-like equation for each isotope)
and the nuclear network equations at each depth mass grid are solved
altogether fully implicit in one scheme. This treatment returns
consistent abundance profiles within the convective region. Thus, the
energy generation rates calculated as a function of position in the
stellar burning region reflects the rapid consumption of protons in
the He-flash convection zone at each position consistently.

  \begin{figure}[t]
    \begin{center}
      \epsfxsize=\textwidth
      \epsfbox{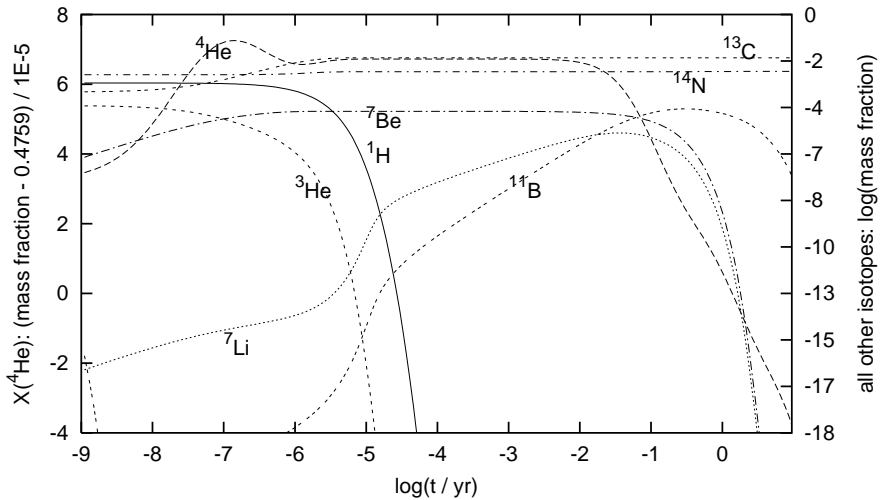}
    \end{center}
    \caption{\label{fig:one-zone}  Time evolution of isotopic abundances
      for constant temperature $T_8=1.5$ and density $\rho =
  1000 \mem{g}/\mem{cm^3}$. The initial abundances resemble the
  AGB envelope composition ($\log X(^3\mem{He}) = -3.9$) with modifications: \hevi, \czw\ and \ose\
  have the intershell abundance of AGB models with overshoot
  (0.47/0.40/0.11) and the initial hydrogen abundance has been chosen
  as $\log X(\mem{H}) = -3$ to account for the dilution effect of the
  mixing (all abundances are given as mass fractions).}
  \end{figure}
\section{Lithium formation during a VLTP}
\label{sec:lithium-VLTP}

The large lithium abundance found in \sak\ must be produced within the
preceding events the VLTP. The lithium can not be inherited from a
previous evolutionary phases. Lithium is destroyed during the giant
evolution and produced only during HBB of massive AGB stars. However,
even then lithium is expected to be destroyed towards the end of
the AGB evolution because the \hedr\ reservoir in the envelope is used
up. Finally, the observations reported by Asplund \etal
(\citeyear{asplund:99a}) show an increase of the lithium abundance
during the born again evolution while the hydrogen abundance is
already very low. That means that a profile of lithium must have
existed below the surface in a region were material has been already
processing hydrogen. In other words, lithium must have been produced
during the actual VLTP. 

The convective hydrogen burning is characterized by the formation of a 
shortlived convection zone which reaches from the location of the
maximum H-burning energy generation up to the outermost layers of the
star (zone 3 in  Fig.\,1). It is separated from the
main He-flash convection zone (zone 2 in  Fig.\,1) by
a tiny radiative layer which is more or less permeable, depending on
the assumptions of overshoot mixing between these two layers. In the
conditions which prevail in zone 3, lithium may be produced by a new
nuclear mechanism of \emph{hot H-deficient $^3$He burning}
\cite{herwig:00f}. \hedr\ from the envelope reacts with \hevi\ as it
enters the hot He-flash 
convection zone and forms \besi. During the normal operation of
the H-shell any \besi\ or possible \lisi\ produced this way has only a
very short lifetime against proton capture. However the situation is different
here because the abundance ratio of protons to
\czw\ is reversed. Hydrogen is the most abundant element in the upper
part of the H-shell. Protons which are injected in the He-flash
convection zone are only a tiny fraction compared to the dominating \czw\
isotopes. The protons are faster captured by \czw\ than \besi\ is formed
from \hedr\ and \hevi. 

\begin{figure}[t]
    \begin{center}
      \epsfxsize=\textwidth
      \epsfbox{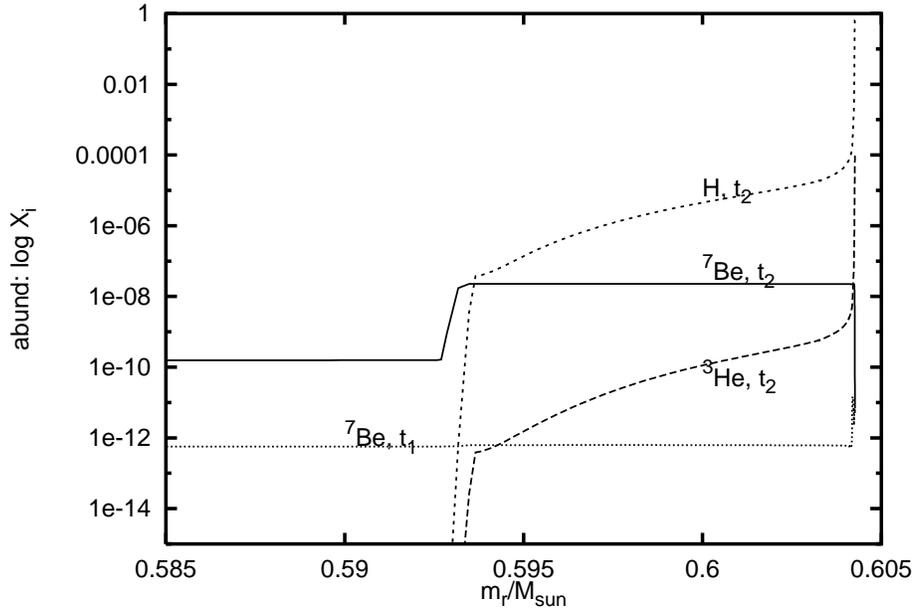}
    \end{center}
    \caption{\label{fig:mod212}  Profiles taken from a coupled convective
      mixing and nuclear burning model sequence of H- and
      \hedr-ingestion into the He-flash convection zone.
$t_\mem{1} = 9.6 \mem{d}$  $t_\mem{2} = 12.7 \mem{d}$ after start of
envelope ingestion. 
}
  \end{figure}
A one-zone nuclear reaction network model
sequence is shown in Fig.\,2. The initial parameters have
been chosen in order to represent the conditions at the bottom of
zone 3 (Fig.\,1). The nuclear network has been
extended by the isotope \bel\ and $\alpha$-capture reactions of \besi,
\lisi\ and \besi\ as well as the \p-capture of \besi. These reactions
have been identified to be important if \hedr\ is ingested into
He-burning. Initially, excess
\hedr\ is burned by the pp-I chain reaction which causes the small
\hevi\ bump at $\log t/\mem{yr} \sim -7$. When the \hedr\ mass
fraction falls below a certain value (here $\sim 10^\mem{-5}$) \besi\
formation starts because the lower reaction rate of the pp-II reaction
is now counterbalanced by the \hevi\ abundance being much larger than
the \hedr\ abundance. At this point the time scale of \hedr\ against
$\alpha$-capture is $\tau_{\alpha}(\hedr) \sim 17\mem{s}$ whereas
$\tau_{^{12}C}(\p) \sim 37\mem{s}$. Thus the protons which are mixed
into the He-flash convection zone together with \hedr\ are much
faster absorbed by 
\czw\ than \besi\ can form. By the time \besi\ has been produced
from the \hedr\ left over after the initial $\hedr + \hedr$ - burning
(pp-I) the hydrogen abundance has been reduced to a negligible
value. 

\besi\ has time to be transformed into \lisi\ by \emi-capture
and the lithium abundance reaches a mass fraction of $\sim 10^{-6}$ at
$\log t/\mem{yr} \apleq -1$. Of course this lithium mass fraction is
based on the initial assumptions for abundances and temperature for
this one-zone  model. It can not be compared directly with the
observed lithium abundance of Sakurai's object. Tests with 
different temperatures have shown that in a hotter regime ($\teight \apgeq
2$) the time scale of $\alpha$-capture on \lisi\ is smaller than that
of \emi\ capture by \besi. No significant lithium abundance can build
up in this case because any \emi-capture is almost immediately
followed by an $\alpha$-capture. For much smaller temperatures
($\teight \apleq 0.8$)  $\tau_{\besi}(\p) < \tau_{\czw}(\p)$ and
protons are not efficiently captured by \czw\ but instead destroy
\besi.

The one-zone model described above gives only a limited approximation
of the turbulent conditions of \hedr-burning in the He-flash
convection zone. I have 
therefore created a preliminary coupled mixing and nucleosynthesis
model of the significant stellar layers. The initial condition,
including the efficiency of convective mixing, has
been chosen according to the VLTP model  by Herwig
\etal (\citeyear{herwig:99c}) just before the He-flash convection zone
reaches the envelope. The time-evolution of the chemical
species has been followed with a diffusion equation for mixing solved
fully implicit and simultaneously with the nuclear network equations.
The upper boundary of the He-flash convection zone has been shifted
outward and into the 
envelope at a rate of  $\mdot = 10^{-3}\msun/\jahre$. Selected
profiles of light elements are shown in  Fig.\,3. During the
first phase of envelope ingestion until $t_\mem{2}=9.6\mem{d}$ any
\besi\ which forms is destroyed immediately at the bottom of the
He-flash convection zone because the split has not yet formed. At
$t_\mem{2}$ this split is introduced and thereafter \besi\ survives
and will eventually form lithium. It appears that this mode of lithium
formation is currently the only conceivable way to explain the large
lithium abundance in Sakurai's object.

  \begin{figure}[t]
    \begin{center}
      \epsfxsize=\textwidth
      \epsfbox{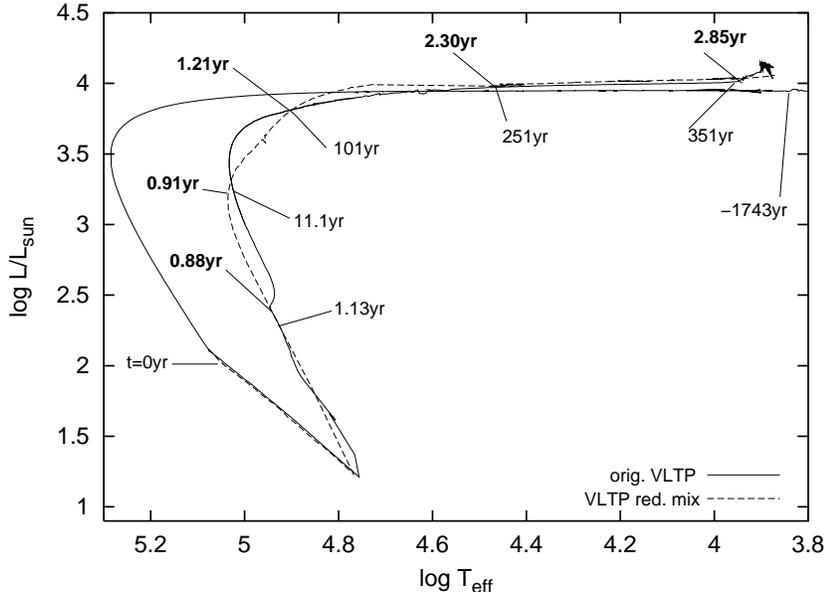}
    \end{center}
    \caption{\label{fig:hrd}  Hertzsprung-Russell diagram of the VLTP
      sequenz of Herwig \etal (1999) (solid line and time
      labels to the bottom and right of track) and a recomputation of
      the flash and subsequent born again evolution under the
      assumption of significantly reduced mixing efficiency of
      elements (dashed line starting at $t=0\jahre$ and bold time
      labels to the left and top of the track). 
}
  \end{figure}
\section{The time scale problem}
\label{sec:time}

Apparently \sak\ has evolved within only a few years from pre-white dwarf
stage to its present state as a luminous and cool giant
\cite{jacoby:98}. This is impossible to reconcile with time scales for the  
born again evolution found in stellar evolution models published so far.

 Two different kinds of models have been constructed. Most
models are those of a \emph{late thermal pulse} evolutionary
channel. Here the thermal pulse occurs while the star is still on the
horizontal crossing from the AGB to the CSPN phase at constant
luminosity. During this first post-AGB phase the H-shell is still
active and prevents the mixing of envelope material into the He-flash
convection zone during the thermal pulse. The star, however, is thrown
back to the AGB regime due to the huge energy release of the He-shell
as a consequence of the thermal pulse. For these LTP models
$\tau_\mem{BA}$ is about $100 \dots 200\jahre$
\cite{bloecker:95b} and thus clearly in disagreement with Sakurai's
object (\kap{sec:ltp}). 

The other group of
computations has simulated the actual VLTP case in which the H-burning
has stopped and the protons in the envelope are mixed down into the
He-flash convection zone where they burn on the convective time scale.
The physical situation of simultaneous burning and convective mixing
on the same time scale at the same location requires a special
numerical treatment. In view of this problem Iben \etal
(\citeyear{iben:83a}) have  presented a model calculation of the VLTP where
they have ignored the nuclear energy released by proton captures
in the He-flash zone in order to follow the hydrostatic structure
evolution. Therefore their model resembles more that of the 
LTP as far as nuclear energy production is concerned. They found $140
< \tau_\mem{BA}/\jahre < 637$ (from their Fig.\,1) which is of the
same order of magnitude than the LTP born again evolution. The next
computation of the VLTP has been presented by Iben and MacDonald
(\citeyear{iben:95b}). In that computation $\tau_\mem{BA} = 17\jahre$,
which is much closer to the born again time scale of \sak, although
still too large by a factor of $3 \dots 4$. Unfortunately the
modelling techniques and physical assumptions are not discussed in
much detail in this short conference paper and the reason for the different
values in $\tau_\mem{BA}$ compared to the earlier model is unknown. 

The model sequence of the VLTP by Herwig \etal (\cite*{herwig:99c})
has been described in \kap{sec:vltp} and as explained the nuclear
energy generation by proton captures in the He-flash convection zone
has been fully taken into account. However, for this sequence we found
$\tau_\mem{BA} \sim 350 \jahre$, somewhat longer than the LTP case
derived from the same AGB starting model. Therefore, this model is not
in agreement with \sak. 

In order to resolve this problem it has been suggested that \sak\
might be very massive. Bl\"ocker (\citeyear{bloecker:95b}) finds that
the evolution from a LTP back to the AGB takes  $290 \jahre$ for the
$0.625\msun$ model and only $53\jahre$ for the $0.836 \msun$ model
sequence. Thus, CSPNe with larger core masses have a shorter
born again phase \cite{bloecker:97}. This has lead Duerbeck \etal
(\citeyear{duerbeck:00}) to the conclusion that \sak\ could be as
massive as $1.0 \msun$.  
Moreover they note that a high mass would require the long distance
scale ($d \simeq 5.4 \kpc$) which is in contradiction to the much
shorter distance of $d \simeq 1.1 \kpc$ obtained from the extinction
method \cite{kimeswenger:98}.
This high mass might resolve the time scale
problem, however, there is good reason against this
solution.\footnote{Jacoby
  \etal\citeyear{jacoby:98} have tentatively estimated distances for
  \sak\ from nebular expansion velocity and diameter together with
  ages from stellar models by Bl\"ocker (\citeyear{bloecker:95b})
  which suggests that more massive models should result in smaller
  distances. However, taking PN ages as the crossing time from the
  Bl\"ocker models implictly assumes that the CSPN experiences a
  LTP. Thus, the Bl\"ocker models, like any LTP models, are not
  applicable to this case. In fact the LTP can statistically occur at
  any time during the crossing in the case of the LTP and thus the PN
  age of a born again AGB star following the LTP has no unique relation with
  the mass. Furthermore,   more massive CSPN of the VLTP variety do not
  necessarily have smaller PN ages. The VLTP occurs typically at $\log
  L/\lsun \apleq 2$. Bl\"ocker (\citeyear{bloecker:95b}) has shown
  that massive pre-white dwarfs actually cool down slower than their
  less massive counterparts. PN ages of massive VLTP born again stars might
  very well be effected by this cooling behaviour. This means in
  summary: (a) PN ages of LTP model stars are not applicaple to \sak\
  and (b) the PN ages of VLTP stars do not necessarily scale down with
  increasing mass.}  

As discussed in \kap{sec:lithium-VLTP}
the high amount of lithium found in \sak\ is not inherited from ap
revious stellar evolution phase but must have been synthesized after
the star left the AGB for the first time. However, any stellar
nucleosynthesis being capable of producing lithium relies on a
reservoir of \hedr. If \sak\ has a lower core mass than $\sim 0.7 \msun$
  then \hedr, built up during the main
sequence evolution, will be preserved in the remaining small envelope
mass of the post-AGB star and the synthesis of lithium via the above
described process is possible. If on the other hand the progenitor of
\sak\ has been as massive as required to resolve the time scale
problem  than the progenitor AGB star would have experienced Hot
Bottom Burning which destroys \hedr\ in order to produce \lisi\ (which
is itself destroyed later on when the \hedr\ reservoir of the envelope
is consumed). In that case it is hard to imagine how lithium could
have been produced during the post-AGB phase without \hedr. A very high
mass of \sak\ is unlikely for another reason. The abundance ratio N/O
in the PN is well below unity \cite{pollacco:99}. The PN
material reflects the envelope composition during the very last
thermal pulses on the AGB. According to recent stellar evolution
calculations by 
Lattanzio \& Forestini (\cite*{lattanzio:99}) AGB stars with the
highest mass have a continuously increasing N/O ratio due to Hot
Bottom Burning and eventually the ratio exceeds unity, e.g.\ for the
$\mzams=6\msun$ model of solar metalicity. Therefore the progenitor of \sak\
was not an AGB with the highest mass. 
Instead the progenitor mass was small enough to avoid
the Hot Bottom Burning phase alltogether and the time scale discrepancy
remains.

\section{The solution of the time scale problem}
\label{sec:solution}
The rapid return of \sak\ to the AGB is another case of the general
problem of why stars become red giants. As discussed in detail by Sugimoto
and Fujimoto (\citeyear{sugimoto:00})  multiple solutions
to stars of similar composition correspond to different topologies of
the non-linear stellar structure 
equations as a boundary value problem. The transition between 
solutions of different topologies can be obtained if
the assumption of thermal equilibrium is relaxed which leads
to the initial value problem of stellar evolution. E.g., in order to
switch from a  dwarf structure to a giant structure the entropy of the
envelope has to be increased.  

In
the VLTP model by Herwig \etal (\citeyear{herwig:99c}) the main
proton-capture energy is released deep in the He-flash convection zone
were the temperature has already been largely increased by the ongoing
He-flash. Thus the entropy increase in this layers by the additional
H-burning luminosity is barely effecting the outermost layers, and
the born again evolution proceeds 
on the time scale determined by the energy release of the
He-flash. Therefore the born again time scale of this VLTP model is
similar to that of the corresponding LTP model.

The position of the peak proton-capture energy release is controlled
by the velocity of convective element mixing and the rate of proton
capture by \czw. The convective velocity defines a time scale of
element mixing $\tau_\mem{mix}$ while the p-capture rate defines a
nuclear time scale $\tau_\mem{nuc}$. The latter decreases with
increasing temperature. As protons are caught by the convective mixing
of the He-flash zone they are transported inward on the time scale
$\tau_\mem{mix}$ while $\tau_\mem{nuc}$ decreases as they reach the
hotter layers. At the position in the He-flash convection zone where
both time scales are equal the peak p-capture energy is released. For
the VLTP model by Herwig \etal (\citeyear{herwig:99c}) this position
is at mass coordinate $m_\mem{r}=0.5950\msun$. 

In order to release the entropy generated from the additional
p-captures into the envelope
the position of this energy release must occur further outwards and
closer to the original lower envelope boundary. This can be achieved
by assuming 
that the efficiency of element mixing in the He-flash convection zone
is \emph{lower}. Then, $\tau_\mem{mix}$ is larger and equals
$\tau_\mem{nuc}$ already at a lower temperature at a larger mass
coordinate. 

I have computed a new VLTP model sequence with a starting model of the
original sequence just before the ingestion of protons into the
He-flash convection zone starts. The diffusion coefficient which is
used to describe the element mixing efficiency according to the
mixing-length theory in a time-dependent way has been decreased by a
factor of $1000$. As a result the peak p-capture energy is released at
$m_\mem{r}=0.6015\msun$ leading to a peak temperature of
$\teight=1.15$. The evolutionary track of this model sequence is shown
in  Fig.\,4
together with the original VLTP sequence and both equipped with time
marks. The evolution back to the AGB is greatly accelerated and
complete within $\sim 3\jahre$. Thus, the evolutionary time scale of
the VLTP model with reduced efficiency of element mixing is in full
agreement with that of \sak. Tests have indicated that these modified
models are as well a suitable host for the \emph{hot H-deficient
  \hedr-burning} described in \kap{sec:lithium-VLTP}.   

This result appears to demonstrate that the convective velocities
obtained from the mixing-length theory are not a particular good
estimate. In fact, Sakurai's object may be used in the future to test
convection theory.

\acknowledgements
This work has been supported in part by the \emph{Deut\-sche
  For\-schungs\-ge\-mein\-schaft, DFG\/} (La\,587/16).  I would like
to thank N.\ Langer and D.A.\ VandenBerg for very useful discussions. 


\begin{thebibliography}{}

\bibitem[\protect\citeauthoryear{{Asplund} et~al.}{1999}]{asplund:99a}
{Asplund}, M., D.~L. {Lambert}, T. {Kipper}, D. {Pollacco}, and M.~D.
  {Shetrone}: 1999, `The rapid evolution of the born-again giant Sakurai's
  object'.
\newblock {\em A\&A} {\bf 343}, 507--518.

\bibitem[\protect\citeauthoryear{Bl\"ocker}{1995}]{bloecker:95b}
Bl\"ocker, T.: 1995, `Stellar evolution of low and intermediate mass stars. II.
  Post-AGB evolution'.
\newblock {\em A\&A} {\bf 299}, 755.

\bibitem[\protect\citeauthoryear{Bl\"ocker}{2000}]{bloecker:00a}
Bl\"ocker, T.: 2000, `Stellar evolution on the AGB and beyound'.
\newblock In: T. Bl\"ocker, R. Waters, and B. Zijlstra (eds.): {\em Low mass
  Wolf-Rayet Stars: origin and evolution}.
\newblock in press.

\bibitem[\protect\citeauthoryear{Bl\"ocker and
  Sch\"onberner}{1997}]{bloecker:97}
Bl\"ocker, T. and D. Sch\"onberner: 1997, `Stellar evolution of low and
  intermediate-mass stars III. An application of evolutionary post-AGB models:
  the variable central star FG Sagittae'.
\newblock {\em A\&A} {\bf 324}, 991.

\bibitem[\protect\citeauthoryear{Duerbeck et~al.}{2000}]{duerbeck:00}
Duerbeck, H.~W., W. Liller, and S.~B. etal.: 2000, `"The rise and fall of V4334
  Sagittarii (Sakurai´s Object)"'.
\newblock {\em AJ} {\bf 119}, 2360--2375.

\bibitem[\protect\citeauthoryear{Herwig}{2000}]{herwig:00a}
Herwig, F.: 2000, `Internal mixing and surface abundances of [WC] CSPN'.
\newblock In: T. Bl\"ocker, R. Waters, and B. Zijlstra (eds.): {\em Low mass
  Wolf-Rayet Stars: origin and evolution}.
\newblock in press, astro-ph/9912353.

\bibitem[\protect\citeauthoryear{Herwig et~al.}{1999}]{herwig:99c}
Herwig, F., T. Bl\"ocker, N. Langer, and T. Driebe: 1999, `On the origin of
  hydrogen-deficient post-AGB stars'.
\newblock {\em A\&A} {\bf 349}, L5.

\bibitem[\protect\citeauthoryear{Herwig and Langer}{2000}]{herwig:00f}
Herwig, F. and N. Langer: 2000, `Convective proton and $^3$He ingestion into
  helium burning: Nucleosynthesis during a Post-AGB thermal pulse'.
\newblock {\em Nucl. Phys. A} p. in press.
\newblock astro-ph/0010120.

\bibitem[\protect\citeauthoryear{Iben et~al.}{1983}]{iben:83a}
Iben, Jr., I., J.~B. Kaler, J.~W. Truran, and A. Renzini: 1983, `On the
  evolution of those nuclei of planetary nebulae that experience a final helium
  shell flash'.
\newblock {\em ApJ} {\bf 264}, 605.

\bibitem[\protect\citeauthoryear{Iben and MacDonald}{1995}]{iben:95b}
Iben, Jr., I. and J. MacDonald: 1995, `The Born Again AGB phenomenon'.
\newblock In: D. Koester and K. Werner (eds.): {\em White Dwarfs}. Heidelberg,
  p.~48, Springer.

\bibitem[\protect\citeauthoryear{Iben and Renzini}{1983}]{iben:83b}
Iben, Jr., I. and A. Renzini: 1983, `Asymptotic Giant Branch evolution and
  beyond'.
\newblock {\em ARA\&A} {\bf 21}, 271.

\bibitem[\protect\citeauthoryear{{Jacoby} et~al.}{1998}]{jacoby:98}
{Jacoby}, G.~H., O. {De Marco}, and D.~G. {Sawyer}: 1998, `The Size and Age of
  Sakurai's Planetary Nebula and the Temperature of it's Central Star'.
\newblock {\em AJ} {\bf 116}, 1367.

\bibitem[\protect\citeauthoryear{Kerber et~al.}{1999}]{kerber:99}
Kerber, F., J. K\"oppen, M. Roth, and S. Trager: 1999, `The hidden past of
  Sakurai's object'.
\newblock {\em A\&A} {\bf 344}, L79.

\bibitem[\protect\citeauthoryear{{Kimeswenger} and
  {Kerber}}{1998}]{kimeswenger:98}
{Kimeswenger}, S. and F. {Kerber}: 1998, `The distance of Sakurai's Object'.
\newblock {\em A\&A} {\bf 330}, L41.

\bibitem[\protect\citeauthoryear{Lattanzio and Forestini}{1999}]{lattanzio:99}
Lattanzio, J. and M. Forestini: 1999, `Nucleosynthesis in AGB stars'.
\newblock In: T.~L. Bertre, A. Lebre, and C. Waelkens (eds.): {\em AGB Stars}.
  p.~31, ASP.

\bibitem[\protect\citeauthoryear{{Pollacco}}{1999}]{pollacco:99}
{Pollacco}, D.: 1999, `The planetary nebula surrounding the final thermal pulse
  object V4334 Sagittarii'.
\newblock {\em MNRAS} {\bf 304}, 127--134.

\bibitem[\protect\citeauthoryear{Sackmann and Boothroyd}{1992}]{sackmann:92}
Sackmann, I.-J. and A.~I. Boothroyd: 1992, `The Creation of Superrich Lithium
  Giants'.
\newblock {\em ApJ} {\bf 392}, L71.

\bibitem[\protect\citeauthoryear{Sch\"onberner}{1979}]{schoenberner:79}
Sch\"onberner, D.: 1979, `Asymptotic Giant Branch Evolution with Steady Mass
  Loss'.
\newblock {\em A\&A} {\bf 79}, 108.

\bibitem[\protect\citeauthoryear{{Sugimoto} and Fujimoto}{2000}]{sugimoto:00}
{Sugimoto}, D. and M.~Y. Fujimoto: 2000, `Why stars become red giants'.
\newblock {\em ApJ} {\bf 538}, 837.

\end{thebibliography}

\end{article}
\end{document}